\documentclass[12pt]{article}
 \usepackage{graphicx}
 \usepackage[cp1251]{inputenc}

\begin{document}
 \noindent {\footnotesize\it
   Astronomy Letters, 2021, Vol. 47, No 8, pp. 534--543.}
 \newcommand{\dif}{\textrm{d}}

 \noindent
 \begin{tabular}{llllllllllllllllllllllllllllllllllllllllllllll}
 & & & & & & & & & & & & & & & & & & & & & & & & & & & & & & & & & & & & & &\\\hline\hline
 \end{tabular}

  \vskip 0.5cm
\centerline{\bf\large Estimation of the Radial Scale Length and Vertical Scale Height}
\centerline{\bf\large of the Galactic Thin Disk from Cepheids}
   \bigskip
   \bigskip
  \centerline
 {V. V. Bobylev \footnote [1]{e-mail: vbobylev@gaoran.ru} and A. T. Bajkova}
   \bigskip

  \centerline{\small\it Pulkovo Astronomical Observatory, Russian Academy of Sciences,}

  \centerline{\small\it Pulkovskoe sh. 65, St. Petersburg, 196140 Russia}
 \bigskip
 \bigskip
 \bigskip

{\bf Abstract}---Up-to-date data on 2214 classical Cepheids have been used. The model of an exponential matter density distribution in the Galactic thin disk has been applied. New estimates of the radial disk scale length $h_R,$ the distance of the Sun from the symmetry plane $z_\odot$, and the vertical disk scale height $h_z$ have been obtained. Based on samples of 1087 Cepheids younger than 120 Myr and 1127 Cepheids older than 120 Myr, we have found $h_R=2.30\pm0.09$ kpc and $h_R=1.96\pm0.12$ kpc, respectively. It has turned out that an unambiguous estimate of $h_R$ cannot be obtained from their overall radial distribution. However, we have constructed the radial distribution of 806 Cepheids of different ages at the time of their birth from which we have found $h_R=2.36\pm0.24$ kpc. We show that $h_z$ depends strongly on the age of stars. For example, from 705 Cepheids younger than 120 Myr with heliocentric distances $r<6$ kpc we have found $z_\odot=-17\pm4$ pc and $h_z=75\pm5$ pc. From 393 Cepheids older than 120 Myr at $r<6$ kpc, we have found
$z_\odot=-39\pm11$ pc and $h_z=131\pm10$ pc.


 \subsection*{INTRODUCTION}
The spatial, kinematic, and elemental structures of the Milky Way stellar disk are complex. It is not yet quite clear how much the disk evolution owes to internal factors and how much to satellite accretion or other external heating factors.

Knowing the specific characteristics of the distribution of matter in the Galactic disk is important for constructing an adequate Galaxy model. At present, the most detailed Galaxy models are the Besancon (Robin and Cr\'ez\'e 1986; Robin et al. 2003) or TRILEGAL
(TRIdimensional modeL of thE GALaxy, Girardi et al. 20051 ) ones, in which various characteristics of the bulge, the thin and thick disks, the interstellar matter, and the stellar and dark matter halos are reflected.

Various model dependences to describe the density distribution in the Galactic disk are known: the model of an exponential density distribution (see, e.g.,
Ojha 2001; Joshi 2007; Bovy et al. 2016), the model of a self-gravitating isothermal disk (Conti and Vacca 1990), and the Gaussian model (see, e.g.,
Maiz-Apell\'aniz 2001; Elias et al. 2006). Other approaches are also applied (Rosslowe and Crowther 2015). These dependences are used to describe the distribution of matter in both radial and vertical directions.

In this paper we are interested in the properties of the Galactic thin disk. Such characteristics of the vertical distribution of matter as the mean $z_\odot$ and the
vertical scale height $h_z$ are determined most easily. Their values can be found directly by analyzing a histogram of the distribution of objects in $z$ coordinate,
i.e., by analyzing an $N-z$ histogram (where $N$ is the number of stars). Objects from a solar neighborhood of even a relatively small size can be used for
this purpose. For example, Piskunov et al. (2006)
found the values of these quantities from open star
clusters (OSCs) selected under the condition of their
completeness within 0.85 kpc of the Sun. The papers
by Elias et al. (2006), Bobylev and Bajkova (2016a, 2016b), and Skowron et al. (2019) are devoted to estimating $z_\odot$ and $h_z$ using various Galactic disk objects.

The radial scale length $h_R$ is more difficult to determine. Highly accurate (with errors less than 10--15\%) heliocentric distances of objects are required to construct a histogram of their distribution in Galactocentric distance $R.$ Until recently, there have been simply no distances to a large number of stars with the required accuracy. Therefore, photometric estimates of the heliocentric distances (with errors 20--25\%) and more complex methods based on the extraction of information from star counts in various
directions from the Sun were used to estimate the radial disk scale length $h_R$. Such methods were applied in Ojha (2001), Reyl\'e and Robin (2001), Girardi et al. (2005), Benjamin et al. (2005), and Bovy et al. (2016).

However, the unique paper by Skowron et al. (2019) has appeared recently, in which highly accurate (with errors $\sim5\%$) heliocentric distances to 2431 classical Cepheids were determined from the period–luminosity relation. These Cepheids are distributed
in wide region of the Galaxy, $3<R<25$~kpc. Therefore,
in this paper, to estimate the radial scale length hR, we are going to apply the direct method based on the analysis of an $N-R$ histogram to these objects.

Thus, the goal of this paper is to redetermine the radial scale length and vertical scale height of the Galactic thin disk. For this purpose, we use data
on classical Cepheids from Skowron et al. (2019). We apply the model of an exponential matter density distribution in both radial and vertical directions.

 \section*{METHODS}
In the case of an exponential matter density distribution in the radial direction (along the distance of a star from the Galactic center $R$), the observed histogram of the distribution of objects is described by the following expression:
 \begin{equation}
  N(R)=N_R \exp\biggl(-{R\over h_R}\biggr),
 \label{exp-R}
 \end{equation}
where $N_R$ is the normalization factor. In the vertical direction (along the $z$ coordinate axis) the analogous expression has a similar form:
 \begin{equation}
  N(z)=N_z \exp\biggl(-{|z-z_\odot|\over h_z}\biggr),
 \label{exp-Z}
 \end{equation}
where the mean $z_\odot$ reflects the Sun’s elevation above the Galactic plane and $N_z$ is the corresponding normalization factor.

In this paper we estimate the parameters $h_R$ and $N_R$ as well as $h_z,$ $z_\odot$, and $N_z$ by the least-squares method. For this purpose, it is convenient to take the logarithm of the right- and left-hand sides of Eqs. (1) and (2); the equations then take a linear form.

Table 1 gives the individual estimates of the radial scale lengths for the Galactic thin and thick disks obtained by various authors in the last 20 years using various star catalogues. A fairly extensive overview of the earlier estimates of these quantities can be found
in Ojha (2001). Note that all the values of $h_R$ listed in the table were obtained by assuming an exponential radial decrease in the star density in accordance with relation (1).

\begin{table}[t]
\caption{Determinations of the radial Galactic disk scale length $h_R$}
\begin{center}
\begin{tabular}{|l|c|c|c|c|}\hline
 $h_R\pm\varepsilon_{h_R},~$kpc & Disk & Comments & References \\\hline
       $2.8  \pm0.3 $ &  thin & star counts from 2MASS data & [1] \\ 
 $3.7^{+0.8}_{-0.5} $ & thick & star counts from 2MASS data & [1] \\ 
 $2.75 \pm0.16$ &  thin &                     1400 M dwarfs & [2] \\ 
 $2.50 \pm0.50$ & thick &                       star counts & [3] \\ 
 $2.80 \pm0.25$ &  thin &        disk in the TRILEGAL model & [4] \\
 $2.53 \pm0.11$ &         &     disk at the Galactic center & [5] \\ 
 $3.9  \pm0.6 $ &         &   star counts from GLIMPSE data & [6] \\ 
 $2.2  \pm0.2 $ &         &    star counts from APOGEE data & [7] \\ 
 $2.67 \pm0.09$ &   thick &       star counts from DES data & [8]) \\\hline
 $2.30\pm0.09$ &  thin & 1087 Cepheids, ${\overline t}\sim 82$~Myr & this paper \\
 $1.96\pm0.12$ &  thin & 1127 Cepheids, ${\overline t}\sim200$~Myr & this paper \\
   \hline
\end{tabular}
\label{t-h-R}
\end{center}
{\def\baselinestretch{1}\normalsize\small
[1] Ojha (2001), [2] Zheng et al. (2001), [3] Reyl\'e and Robin (2001), [4] Groenewegen et al. (2002), [5] Robin et al. (2003), [6] Benjamin et al. (2005), [7] Bovy et al. (2016), 
[8] Pieres et al. (2020).
 }
\end{table}

The greater is the value of the parameter $h$ in Eqs. (1) and (2), the slower is the decrease in the matter density and the more extended is the structure. The values of $h_R$ collected in Table 1 correspond to the so-called short disk scale length; the values
exceeding 4~kpc correspond to the long scale length. The value of $h_R=4.1\pm0.4$~kpc found by de Jong et al. (2010) for the thick disk or $h_R=5.5\pm1$~kpc found by van der Kruit (1986) serve as examples.

The Galactic disk is usually divided into the thin and thick ones. They are assumed to have different kinematic and evolutionary properties, although this division is often fairly formal. Some authors dispute the necessity of dividing the disk into the thin and
thick ones by assuming that the disk is more complex and consists of a mixture of several mono-abundance populations (MAPs) (Bovy et al. 2012, 2016; Rix and Bovy 2013). Nevertheless, the disk type to which the cited authors themselves assign their $h_R$ estimate is specified in the second column of Table 1.

For example, Ojha (2001) found hR using data on stars from the near-infrared Two Micron All Sky Survey (2MASS, Skrutskie et al. 2006). The star counts were performed in seven fields located in different parts of the sky. Mostly G-, K-, and M-type stars were the objects of these counts. The stars were separated into members of the thin and thick disks by $J-K_s$ color.

 \begin{table}[t]                                     
 \caption[]{\small
 The parameters of the vertical exponential distribution $z_\odot$ and $h_z$ found by various authors   }
  \begin{center}  \label{t-h-z}   
  \begin{tabular}{|c|c|r|c|c|c|}\hline
  $z_\odot\pm\varepsilon_{z_\odot},$~pc &  $h_z\pm\varepsilon_{h_z},$~pc & $N_\star$ & Objects & References \\\hline
 $~-5.7\pm0.5$  & $26.5\pm0.7$ & 639 & masers, $R<R_0$~kpc &[1]\\
 $~-13.5\pm2.6$ & $30.7\pm5.9$ &  19 &           magnetars &[2]\\
 $~-16\pm5$     & $34.0\pm2.0$ & 553 & OB stars from the Hipparcos catalogue &[3]\\
 $~-10\pm4$     & $51.3\pm3.7$ & 148 &         Wolf–Rayet stars, $r<4.5$~kpc &[4]\\
 $~-22\pm4$     & $56.0\pm3.0$ & 259 &                    OSCs, $r<0.85$~kpc &[5]\\
 $~-23\pm2$     & $70.2\pm2.4$ & 246 &  Cepheids, ${\overline t}\sim 75$~Myr &[4]\\
 $~-14.5\pm3.0$ & $73.5\pm3.2$ & 619 &  Cepheids, ${\overline t}\sim 94$~Myr &[6]\\
 $~-24\pm2$     & $83.8\pm2.4$ & 250 &  Cepheids, ${\overline t}\sim138$~Myr &[4]\\\hline
 $~-17\pm4 $&$ 75\pm5 $ &705& Cepheids, ${\overline t}\sim 83$~Myr & this paper\\
 $~-39\pm11$&$131\pm10$ &393& Cepheids, ${\overline t}\sim182$~Myr & this paper\\
   \hline
 \end{tabular}\end{center}
 {\small
[1] Bobylev and Bajkova (2016b), [2] Olausen and Kaspi (2014), [3] Elias et al. (2006), [4] Bobylev and Bajkova (2016a), [5] Piskunov et al. (2006), [6] Skowron et al. (2019).
   }
 \end{table}

Zheng et al. (2001) used data on 1400 M-type dwarfs that were observed with the Hubble Space
Telescope in more than 300 sky fields. To perform the star counts, Reyl\'e and Robin (2001) used data from various catalogues. Robin et al. (2003) described
the results of determining $h_R$ obtained by analyzing
DENIS (DEep Near-Infrared southern sky Survey,
Epchtein et al. 1997) data, which they attributed
to the central part of the Galaxy. To estimate the
radial scale length of the thin disk, Groenewegen et al. (2002) selected stars from the ESO Imaging Survey (Groenewegen et al. 2002) under the following constraints: $1.3^m<B-V<2^m$,
$20^m<B<22^m$, $1.8^m<V-I<4^m$ and $18^m<I<20^m$.

Benjamin et al. (2005) used data from the GLIMPSE (Galactic Legacy Mid-Plane Survey Extraordinaire, Benjamin et al. 2003) catalogue to estimate the parameters of the central Galactic bar and the radial disk scale length. The observations were performed from
the SPITZER Space Telescope (Werner et al. 2004) in four 3.6~$\mu$m, 4.5~$\mu$m, 5.8~$\mu$m
and 8.0~$\mu$m. photometric bands. In the opinion of these authors, most of the selected
stars are M- and K-type giants. Bovy et al. (2016)
used red-clump giants from the APOGEE (Apache Point Observatory Galactic Evolution Experiment, Eisenstein et al. 2011) infrared spectroscopic survey.

Pieres et al. (2020) used data from the DES (Dark Energy Survey, Abbott et al. 2005, 2018) wide-field photometric survey in the $griz$ bands. TRILEGAL was taken as a model of the Galaxy. The main goal of these authors was to determine the parameters of the
thick disk and the stellar halo; therefore, $h_R,$ $z_\odot$ and $h_z$ referring to the thin disk were fixed with the values adopted in TRILEGAL.

Table 2 gives the individual estimates of the parameters of the vertical exponential distribution $z_\odot$ and $h_z$ found by various authors fromyoung thin-disk objects.

Bobylev and Bajkova (2016a) estimated $z_\odot$ and $h_z$ using data on HII regions, OB associations, Wolf–Rayet stars, classical Cepheids, and masers with
measured trigonometric parallaxes. The distribution
of masers was shown to be severely distorted by the influence of the Local arm.

Bobylev and Bajkova (2016b) analyzed HII regions, giant molecular clouds, and methanol masers, which are distributed over the entire Galaxy. The distances to these objects were determined by the kinematic method. The objects belonging to the Local arm were excluded.

Olausen and Kaspi (2014) used only 19 magnetars.
These magnetars are distributed over the entire
Galaxy. The value of $h_z$ found from the magnetars
is close to those found from the youngest Galactic
objects—masers and OB stars.

Elias et al. (2006) developed a three-dimensional spatial classification method to estimate the structure of the Gould Belt and the Galactic disk. This method
was applied to a sample of OB stars within 1 kpc
of the Sun. As a result, they estimated $z_\odot$ and $h_z$ referring to both the Gould Belt structure (we do not provide them) and the Galactic thin disk.

Piskunov et al. (2006) estimated $z_\odot$ and $h_z$ based on a sample of open star clusters from the ASCC-2.5 catalogue. They took objects within 850 pc of the Sun, where the catalogue is complete.

To estimate $z_\odot$ and $h_z$ from classical Cepheids, Skowron et al. (2019) did not divide the sample by stellar age, but used the following constraints on the latitude $b$ and the distance $R: |b|\leq4^\circ$ and $R<R_0.$ In addition, they took into account the warping of the
Galactic disk.

 \section*{DATA}
In this paper we use data on classical Cepheids from Skowron et al. (2019) and Mr\'oz et al. (2019). The catalogue of Skowron et al. (2019) contains distance, age, pulsation period estimates and photometric data for 2431 Cepheids. These Cepheids
were observed within the fourth stage of the OGLE
(Optical Gravitational Lensing Experiment, Udalski et al. 2015) program. Their apparent magnitudes lie in the range from $I=11^m$ to $I=18^m$. Therefore, a deficit of bright and well-studied Cepheids known from earlier observations is observed here.

 \begin{figure} [t] {\begin{center}
  \includegraphics[width=140mm]{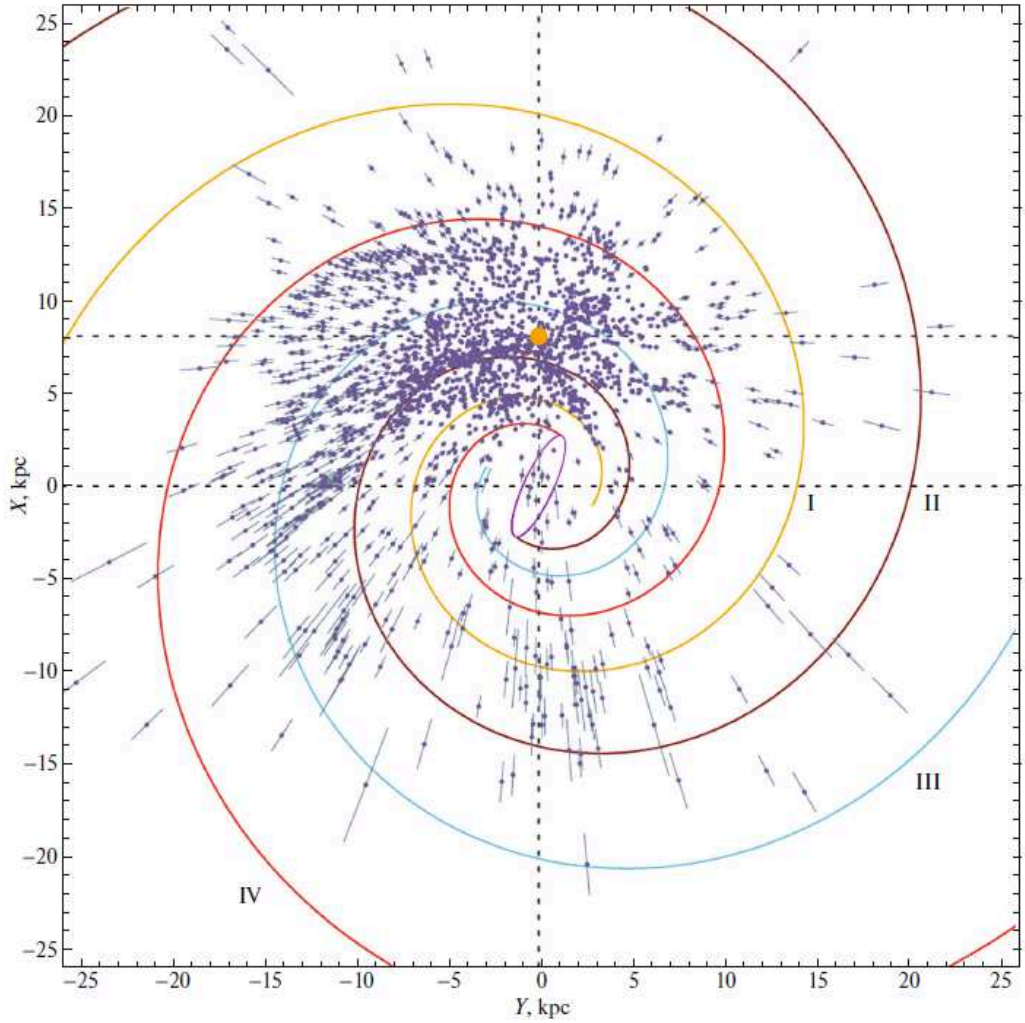}
 \caption{
Positions of 2214 Cepheids in projection onto the Galactic $XY$ plane; the Sun is marked by the orange circle (for
details, see the text).
 }
 \label{f1-XY-00}
 \end{center} }
 \end{figure}
 \begin{figure} [t] {\begin{center}
  \includegraphics[width=150mm]{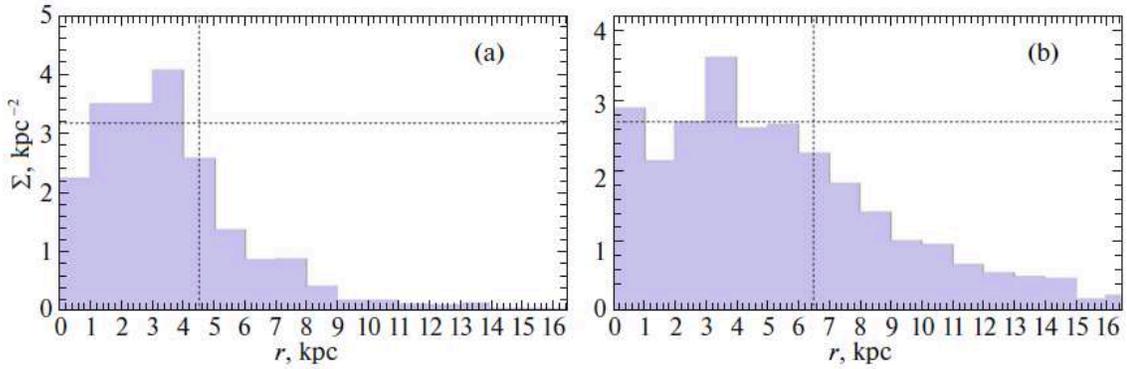}
 \caption{
Surface densities of the samples of the youngest (a) and oldest (b) Cepheids; the mean surface densities and completeness boundaries are indicated.
 }
 \label{f1-common-00}
 \end{center} }
 \end{figure}
 \begin{figure} [t] {\begin{center}
  \includegraphics[width=150mm]{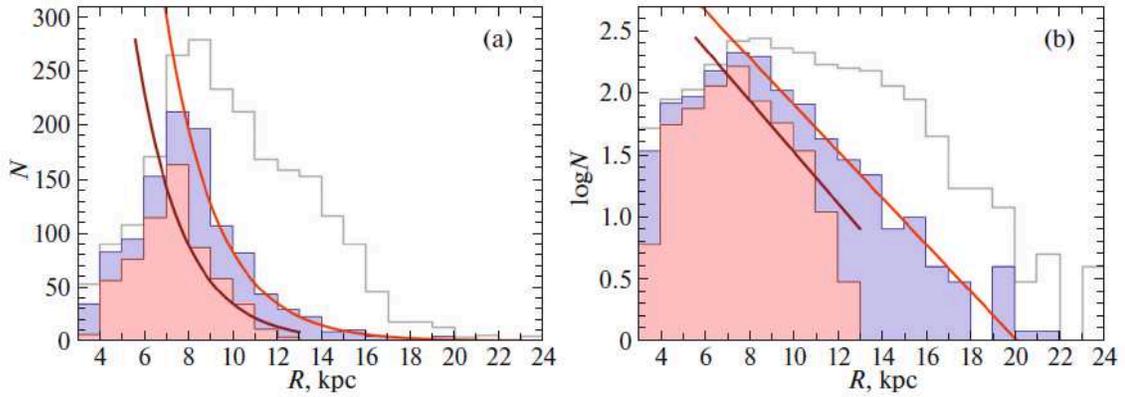}
 \caption{
Histograms of the radial distribution of Cepheids on linear (a) and logarithmic (b) scales; the gray contour, the blue shading, and the red shading indicate the histograms constructed from all 2214 Cepheids, from the Cepheids younger than 120 Myr, and from the Cepheids younger than 120 Myr selected within 5 kpc of the Sun, respectively (for details, see the text).
 }
 \label{f-R-1}
 \end{center} }
 \end{figure}
 \begin{figure} [t] {\begin{center}
  \includegraphics[width=150mm]{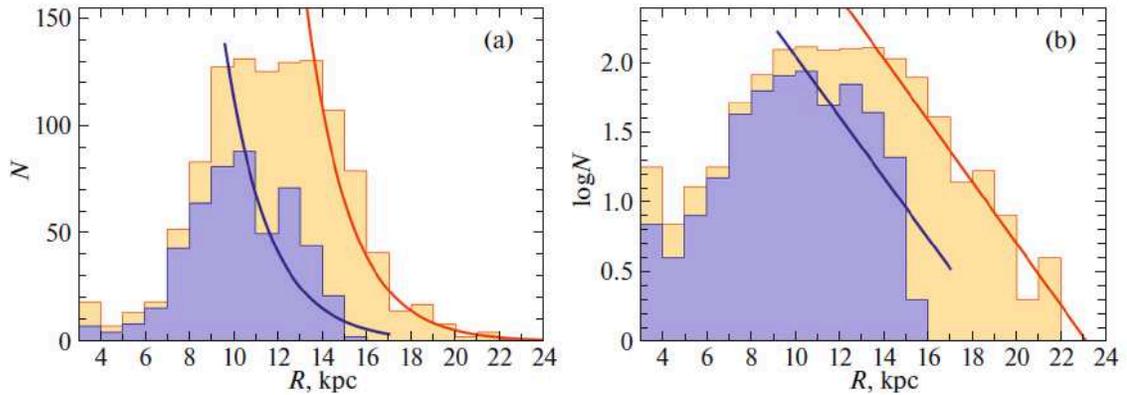}
 \caption{
The orange shading indicates a histogram of the radial distribution of the sample of Cepheids older than 120 Myr on
linear (a) and logarithmic (b) scales; the blue shading indicates a histogram of the radial distribution of the sample of Cepheids
older than 120 Myr selected within 7 kpc of the Sun (for details, see the text).
 }
 \label{f-R-old-1}
 \end{center} }
 \end{figure}
 \begin{figure} [t] {\begin{center}
  \includegraphics[width=150mm]{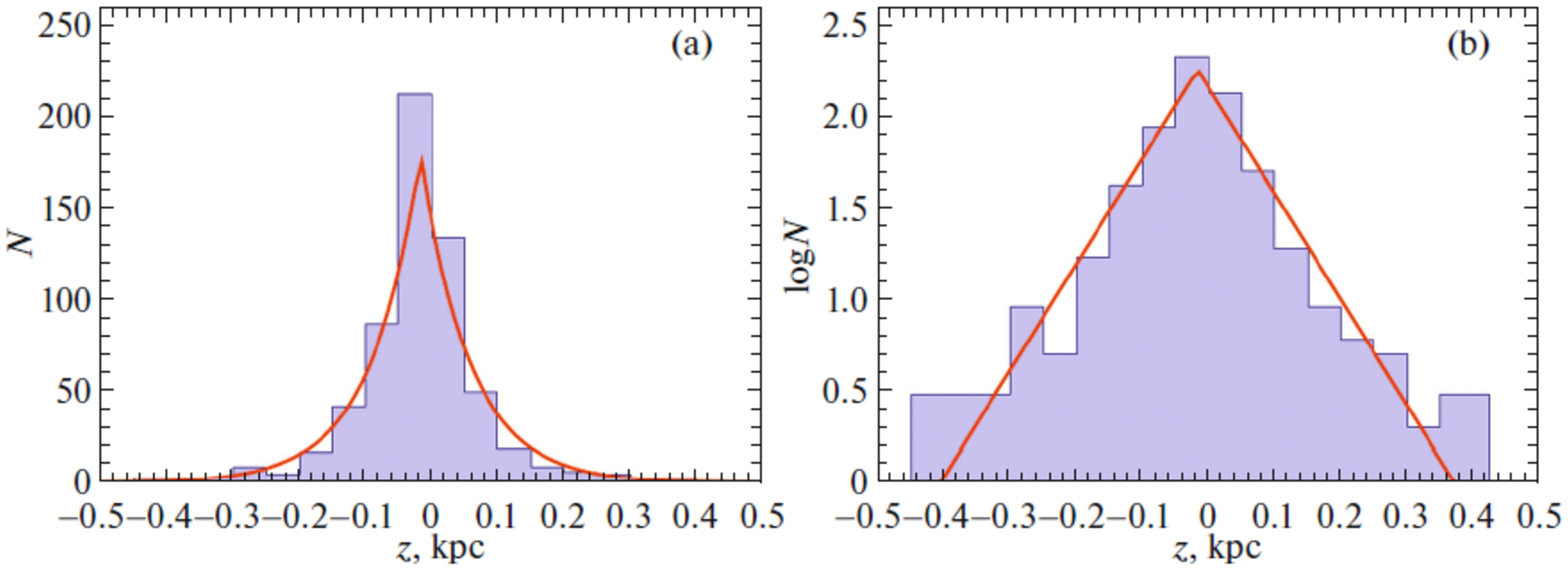}
 \caption{
Histogram of the vertical distribution of the sample of 705 Cepheids younger than 120 Myr with heliocentric distances $r<6$~kpc on linear (a) and logarithmic (b) scales (for details, see the text).
 }
 \label{f-z-1}
 \end{center} }
 \end{figure}
 \begin{figure} [t] {\begin{center}
  \includegraphics[width=150mm]{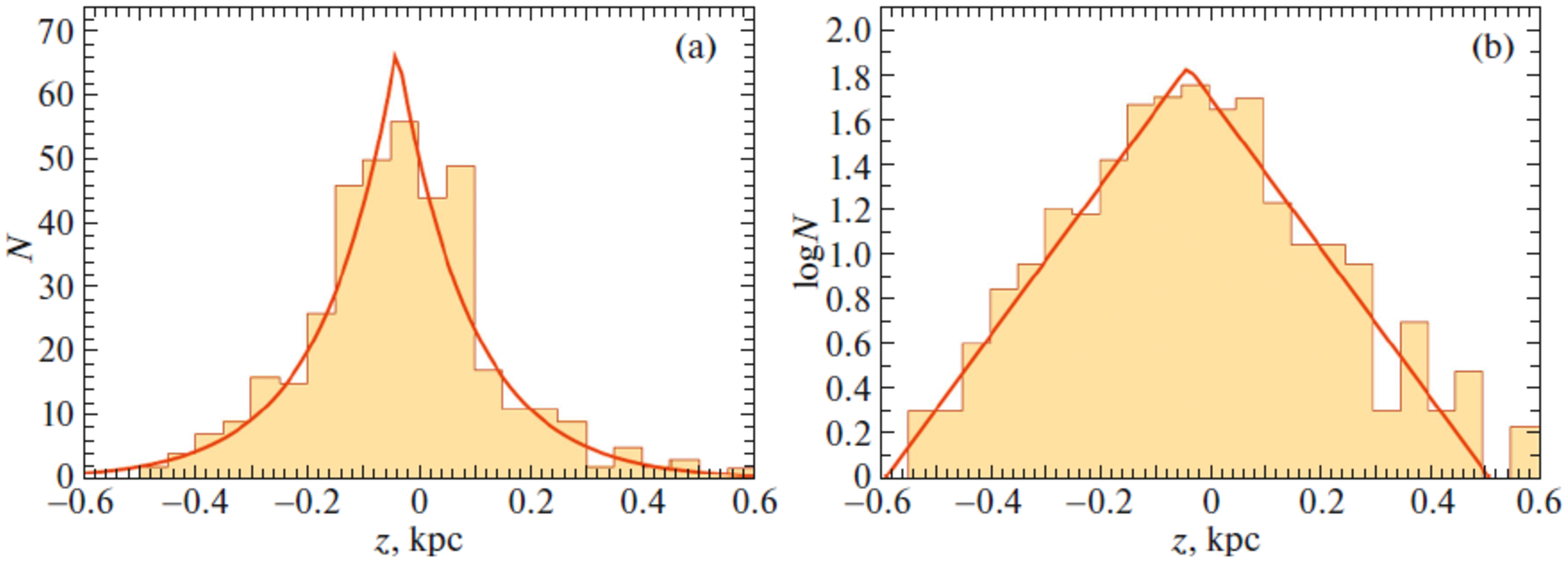}
 \caption{
Histogram of the vertical distribution of the sample of 393 Cepheids older than 120 Myr with heliocentric distances $r<6$ kpc on linear (a) and logarithmic (b) scales (for details, see the text).
 }
 \label{f-z-OLD-1}
 \end{center} }
 \end{figure}

The distances to the stars, $r,$ were calculated by Skowron et al. (2019) based on the period-luminosity relation found by Wang et al. (2018) from
the light curves of Cepheids in the mid-infrared,
where the interstellar extinction is much lower than
that in the optical one. Note that the heliocentric distances in the catalogue of Skowron et al. (2019) were estimated for 2214 Cepheids.

The Cepheid ages in Skowron et al. (2019) were estimated by the technique developed by Anderson et al. (2016). Such characteristics as the stellar rotation periods and metallicity indices were taken into account.

The catalogue of Mr\'oz et al. (2019) is interesting in that kinematic characteristics are given for 832 classical Cepheids from the list of Skowron et al. (2019). These include the proper motion components copied from the Gaia DR2 catalogue
(Lindegren et al. 2018) and the line-of-sight velocities
available for all 832 stars. Using these data, Mr\'oz et al. (2019) constructed the Galactic rotation curve in the range of distances R 4--20 kpc. In Ablimit et al. (2020) the data on these Cepheids were included in the total sample of 3500 classical
Cepheids, which was used to refine the parameters of the Galactic gravitational potential and to obtain a new virial estimate of the Galactic mass. Using the data of Mr\'oz et al. (2019) and Skowron et al. (2019), Bobylev et al. (2021) showed that even old Cepheids
retain kinematic memory of their birthplace.

In this paper we call the sample of Cepheids from the catalogue of Mr\'oz et al. (2019) the kinematic one. For each star from this sample Bobylev et al. (2021) traced the Galactic orbits constructed into the past in accordance with the individual Cepheid age. An
axisymmetric three-component (a bulge, a disk, and a halo) gravitational potential of the Galaxy was used for this purpose. The kinematic sample contains 806 Cepheids. Among them there are 297 stars older than 120 Myr. The mean age of the sample is 115 Myr.

The questions of sample completeness are of great importance for obtaining unbiased estimates of the distribution being analyzed.

Figure 1 presents the distribution of all the Cepheids from the catalogue of Skowron et al. (2019) for which distance estimates are available on the Galactic $XY$ plane. There are a total of 2214 such stars. A four-armed spiral pattern with a pitch angle of $-13^\circ$
(Bobylev and Bajkova 2014) is shown. The roman numerals in the figure number the following spiral arm segments: I~--- Scutum, II~--- Carina--Sagittarius, III~--- Perseus, and IV~--- the Outer arm; the central Galactic bar is shown schematically.

As can be seen from the figure, some of the Cepheids have a strong concentration to the Carina--Sagittarius spiral arm segment. These are mostly Cepheids with ages from 80 to 130 Myr. There are 753 such stars in our sample.

Figure 2 presents the surface brightness distribution for two samples of Cepheids---the youngest and the oldest ones; 983 Cepheids older than 130 Myr (with a mean age of 211 Myr) are distributed most uniformly, and their surface density $\sum$ as a function
of distance $r$ is presented in Fig. 2b. The completeness boundary for the sample of Cepheids older than 130 Myr is 6.5 kpc. We do not provide an analogous figure for the Cepheids of the middle age (from 80 to 130 Myr), because almost all of them are concentrated to the Carina–Sagittarius spiral arm segment, their density drops rapidly with $r.$

The youngest Cepheids have a concentration to the Scutum arm; nevertheless, their distribution on the $XY$ plane is fairly uniform. The behavior of their surface density
$\sum$ as a function of $r$ is presented in Fig. 2a, where the completeness boundary is 4.5 kpc.

On the whole, we can conclude that within 5--6 kpc the sample of Cepheids of various ages is complete. This allows us to estimate the characteristics of Cepheids in such a characteristic range both in the solar neighborhood, for example, to determine
the parameters of the vertical distribution, and in the radial direction away from the Galactic center, i.e., to study the parameters of the radial distribution.

 \section*{RESULTS AND DISCUSSION}
Figure 3 presents the radial distribution of two samples of Cepheids. The first sample includes all of the 2214 Cepheids with various ages for which the heliocentric distances were determined. A histogram of this sample is indicated in the figure by the
gray contour without shading. We see that in the $R$
range 8--22 kpc there is an incomprehensible, at first
glance, hump. Fitting dependences of the form (1)
gives $h_R$ in a very wide range, from 3.3 (in the range
of distances $R$ 8--22 kpc) to 5.5 kpc (in the range
of distances $R$ 8--15 kpc, the very flat dependence in Fig. 3b). These values agree poorly both between themselves and with the known estimates for the thin disk.

Bobylev et al. (2021) also noticed (see Fig. 7 there) that the young and old Cepheids from the catalogue of Mr\'oz et al. (2019) have greatly differing observed
radial distributions. More specifically, the youngest
ones are concentrated within the solar circle, while
the oldest ones are concentrated far beyond the solar circle. Therefore, we decided to divide the Cepheids approximately equally into two samples, depending on the age.

The second sample in Fig. 3 includes 1087 Cepheids younger than 120 Myr. Theirmean age is ${\overline t}\sim82$~Myr. A histogram of this sample is shown in Fig. 3 with
blue shading. For this sample of Cepheids we fitted
an exponential of the form (1) with the radial scale
length $h_R=2.30\pm0.09$~kpc that was found by the least-squares method.

Finally, the third histogram in Fig. 3 was constructed from the Cepheids younger than 120 Myr that were selected within 5 kpc of the Sun. This
sample of stars satisfies the completeness property
and contains 605 Cepheids with a mean age of 83 Myr. The exponential of the form (1) fitted for this histogram has a radial scale length $h_R=2.09\pm0.20$~kpc.

Figure 4 presents the radial distribution of the sample of Cepheids older than 120 Myr. The sample consists of 1127 Cepheids with a mean age ${\overline t}\sim200$~Myr. For this sample of Cepheids we fitted an exponential of the form (1) with a radial scale length $h_R=1.96\pm0.12$~kpc.

The second histogram in Fig. 4 indicated by the blue shading was constructed from the Cepheids older than 120 Myr that were selected within 7 kpc
of the Sun. This sample of stars satisfies the completeness
property and contains 496 Cepheids with a mean age of 186 Myr. The exponential of the
form (1) fitted for this histogram has a radial scale length $h_R=1.99\pm0.29$~kpc. The old Cepheids are greatly shifted toward the Galactic anticenter and, therefore, the histogram of the second sample has a much greater dispersion, $h_R$ is determined with a much larger error. However, the values of $h_R$ themselves derived from the two samples of relatively
old Cepheids turned out to be close. As can be seen from Fig. 4b, the lines on the graph run almost parallel to each other.

 \begin{figure} [t] {\begin{center}
  \includegraphics[width=150mm]{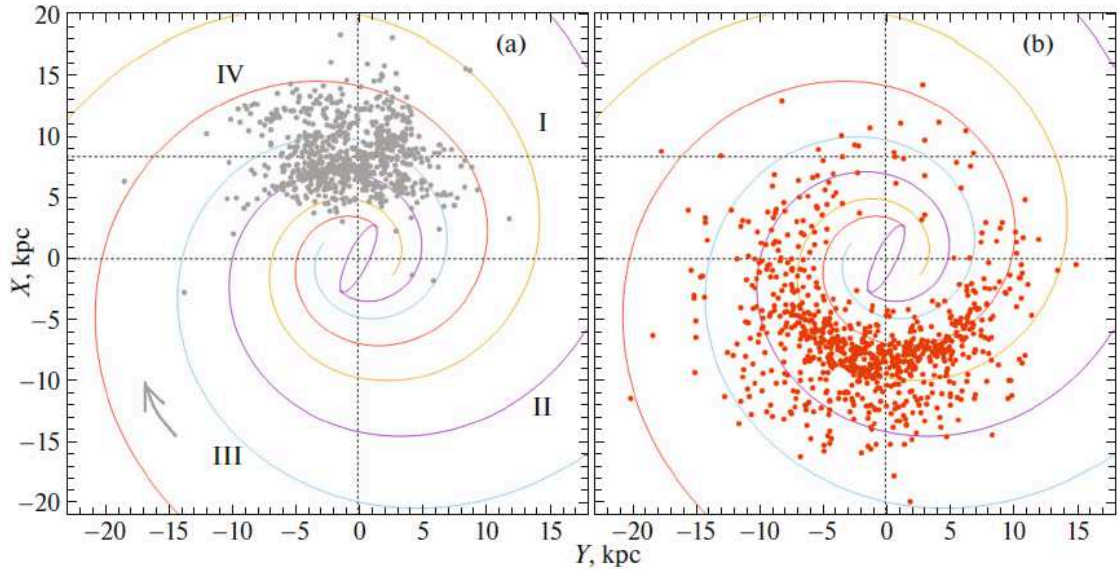}
 \caption{
(a) Present-day positions of the kinematic sample of 806 Cepheids on the Galactic XY plane and (b) their positions in the past calculated for the time of their birth (for details, see the text).
 }
 \label{f-XY-all}
 \end{center} }
 \end{figure}
 \begin{figure} [t] {\begin{center}
  \includegraphics[width=150mm]{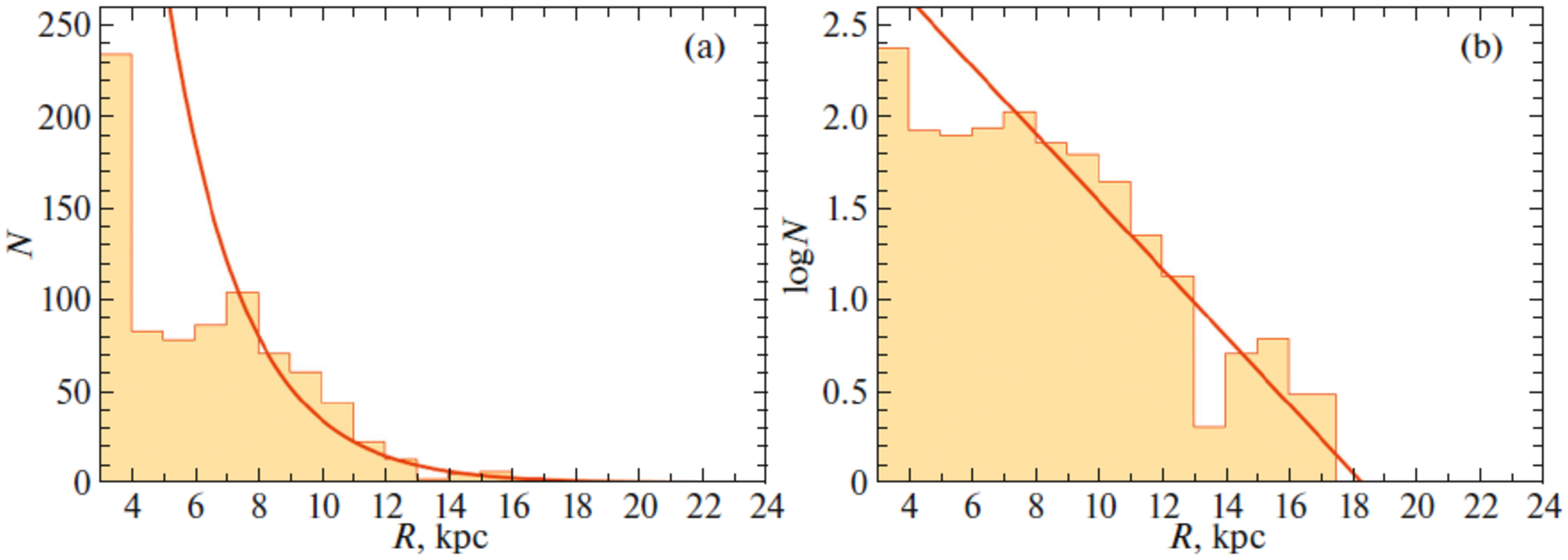}
 \caption{
Histogram of the radial distribution of the kinematic sample of 806 Cepheids whose positions were calculated for the
time of their birth on linear (a) and logarithmic (b) scales (for details, see the text).
 }
 \label{f-R-ALL}
 \end{center} }
 \end{figure}

It can be seen from Figs. 3 and 4 that separately the distributions of the samples of relatively young and relatively old Cepheids satisfy well the model of an
exponential decrease in the radial density. However,
it is better not to use the distribution of the entire
sample (the gray contour in Fig. 3) to determine the radial scale length, because it is the sum of two greatly differing distributions.

Figure 5 presents the vertical distribution of the sample of 705 Cepheids younger than 120 Myr with heliocentric distances $r<6$~kpc and a mean age ${\overline t}\sim
83$~Myr. We fitted an exponential of the form (2) with $z_\odot=-17\pm4$~pc and $h_z=75\pm5$~pc. Having analyzed several individual $z_\odot$ determinations, Bobylev and Bajkova (2016a) calculated the mean $z_\odot=-16\pm2$~pc. Thus, we have good agreement of $z_\odot$ found from young Cepheids with the estimates of other authors (see Table 2).

Figure 6 presents the vertical distribution of the sample of 393 Cepheids older than 120 Myr with heliocentric distances $r<6$~kpc and a mean age ${\overline t}\sim182$~Myr. We fitted an exponential of the form (2) with $z_\odot=-39\pm11$~pc and $h_z=131\pm10$~pc. We see that the Cepheids of this sample are more ``heated'' than
the young Cepheids. At the same time, the vertical
scale height found from them allows these Cepheids
to be attributed to the Galactic thin disk, because a
much larger $h_z\sim900$~kpc is typical for the thick disk (see, e.g., Pieres et al. 2020).

It is well known that there is a close correlation of the spatial distribution of Cepheids and their kinematics with the large-scale warping of the Galactic disk (Berdnikov 1987; Bobylev 2013). The doubling of $h_z$ for Cepheids in $\sim100$~Myr is probably related to
the influence of the disk warping. The disk warping is most likely caused by the Large Magellanic Cloud. Therefore, we can say that an external factor is responsible for the heating of Cepheids.

Figure 7 presents the positions of the kinematic sample of 806 Cepheids in projection onto the Galactic $XY$ plane. The Sun here has coordinates $(X,Y)=(8.3,0)$~kpc. The arrow in the lower left corner in Fig. 7a indicates the direction of Galactic rotation. We can conclude from Figs. 3, 4, and 7 that, for some reasons, quite a few relatively old Cepheids
from the list of Skowron et al. (2019) turned out to be concentrated in the Galactic anticenter region. However, at the time of their birth these stars, along with all of the Cepheids being studied here, were distributed rather uniformly over a large region of the Galaxy.

Figure 8 presents a histogram of the radial distribution of the kinematic sample of Cepheids. The positions of each of these stars in the Galaxy (Fig. 7) were calculated for the time of their birth. In contrast to the overall histogram constructed from Cepheids of
various ages (the gray contour in Fig. 3), in this figure there is no hump near $R\sim14$~kpc, whose presence did not allow $h_R$ to be unambiguously estimated previously.
The radial scale length $h_R=2.36\pm0.24$~kpc found from this sample is in excellent agreement with the results of a separate analysis of the present-day positions of Cepheids (see Table 1).

 \section*{CONCLUSIONS}
To estimate the radial scale length hR and vertical
scale height hz of the Galactic thin disk, we used
a huge sample of classical Cepheids from Skowron
et al. (2019). Quite recently, astronomers had measured
characteristics of $\sim$600 classical Cepheids at
their disposal. The catalogue of Skowron et al. (2019)
contains distance, age, pulsation period estimates
and photometric data for 2431 Cepheids. In this
paper we used 2214 Cepheids from this catalogue
with measured distances.

We estimated $h_R$ and $h_z$ by assuming an exponential star density distribution in both radial and vertical directions

From a sample of 1087 Cepheids younger than 120 Myr we found the radial scale length $h_R=2.30\pm0.09$~kpc. From 1127 Cepheids older than 120 Myr we found $h_R=1.96\pm0.12$~kpc. We analyzed the Cepheids for sample completeness. The distribution
of younger Cepheids was shown to satisfy the completeness
property within about 5 kpc of the Sun.
From such a sample containing 605 Cepheids with
a mean age of 83~Myr we found the radial scale length
$h_R=2.09\pm0.20$~kpc. The completeness boundary
for a sample of older Cepheids is 6.5 kpc. For a sample of such Cepheids (496 with a mean age of 186~Myr) we found the radial scale length $h_R=1.99\pm0.29$~kpc.

We can conclude that there is good agreement with the results obtained from the samples of Cepheids without applying any constraints on the sample radius. Therefore, we prefer to use the $h_R$ estimates with smaller errors.

We showed that an unambiguous estimate cannot be obtained from the total sample of these Cepheids. On the other hand, it was interesting to analyze the
radial distribution of all these Cepheids at the time of
their birth. To perform this task, we used a sample of
806 Cepheids for which it was possible to construct
their Galactic orbits. It turned out that $h_R=2.36\pm0.24$~kpc is determined with confidence by analyzing their radial distribution in the past constructed for
the birth time of each star. We see that all three $h_R$ estimates coincide within the error limits. Thus, no evolution of this parameter is observed on a time scale $\sim$100 Myr. This cannot be said about $h_z,$ whose value approximately doubles on such a time scale (there is disk heating).

Constraints on the age, size, and location of the
region being analyzed are required to estimate the
vertical scale height due to the influence of various
factors. Therefore, not all of the Cepheids were involved
in our analysis here. Furthermore, the vertical
scale height depends strongly on the age of the
sample stars. Therefore, the sample was divided into
two parts by age with a boundary of 120 Myr. From
705 Cepheids younger than 120 Myr with heliocentric distances $r<6$~kpc we found $z_\odot=-17\pm4$~pc and $h_z=75\pm5$~pc, while from a sample of 393 Cepheids older than 120 Myr and with $r<6$~kpc we found $z_\odot=-39\pm11$~pc and $h_z=131\pm10$~pc.

\bigskip{\bf ACKNOWLEDGMENTS}

We are grateful to the referee for the useful remarks that contributed to an improvement of the paper.

\bigskip \bigskip\medskip{\bf REFERENCES}{\small

1. T. Abbott, G. Aldering, J. Annis, M. Barlow, C. Bebek, B. Bigelow, C. Beldica, R. Bernstein, et al. (DES Collab.), arXiv: astro-ph/0510346 (2005).

2. T. Abbott, F. B. Abdalla, S. Allam, A. Amara, J. Annis,
J. Asorey, S. Avila, O. Ballester, et al. (DES
Collab., Astrophys. J. Suppl. Ser. 239, 18 (2018).

3. I. Ablimit, G. Zhao, C. Flynn, and S. A. Bird, Astrophys.
J. 895, L12 (2020).

4. R. I. Anderson, H. Saio, S. Ekstr\"om, C. Georgy, and
G. Meynet, Astron. Astrophys. 591, A8 (2016).

5. R. A. Benjamin, E. Churchwell, B. L. Babler,
T. M. Bania, D. P. Clemens, M. Cohen, J. M. Dickey,
R. Indebetouw, et al., Publ. Astron. Soc. Pacif. 115,
953 (2003).

6. R. A. Benjamin, E. Churchwell, B. L. Babler, R. Indebetouw, M. R. Meade, B. A. Whitney, C. Watson, M. G. Wolfire, et al., Astrophys. J. 630, L149 (2005).

7. L. N. Berdnikov, Sov. Astron. Lett. 13, 45 (1987).

8. V. V. Bobylev, Astron. Lett. 39, 819 (2013).

9. V. V. Bobylev and A. T. Bajkova, Mon. Not. R. Astron.
Soc. 437, 1549 (2014).

10. V. V. Bobylev and A. T. Bajkova, Astron. Lett. 42, 1
(2016a).

11. V. V. Bobylev and A. T. Bajkova, Astron. Lett. 42, 182
(2016b).

12. V. V. Bobylev, A. T. Bajkova, A. S. Rastorguev, and
M. V. Zabolotskikh, Mon. Not. R. Astron. Soc. 502,
4377 (2021).

13. J. Bovy, H.-W. Rix, C. Liu, D. W. Hogg, T. C. Beers,
Y. S. Lee, et al., Astrophys. J. 753, 148 (2012).

14. J. Bovy, H.-W. Rix, E. F. Schlafly, D. L. Nidever,
J. A. Holtzman, M. Shetrone, T. C. Beers, et al.,
Astrophys. J. 823, 30 (2016).

15. P. S. Conti and W. D. Vacca, Astron. J. 100, 431
(1990).

16. D. J. Eisenstein, D. H. Weinberg, E. Agol, H. Aihara,
C. A. Prieto, S. F. Anderson, J. A. Arns, E. Aubourg,
et al., Astron. J. 142, 72 (2011).

17. F. Elias, J. Cabrero-Ca\~no, and E. J. Alfaro, Astron. J. 131, 2700 (2006).

18. N. Epchtein, B. de Batz, L. Capoani, L. Chevallier, E. Copet, P. Fouqu\'e, P. Lacombe, T. le Bertre, et al., Messenger 87, 27 (1997).

19. L. Girardi, M. A. T. Groenewegen, E. Hatziminaoglou, and L. da Costa, Astron. Astrophys. 436, 895, (2005).

20. M. A. T. Groenewegen, L. Girardi, E. Hatziminaoglou, C. Benoist, L. F. Olsen, L. da Costa, S. Arnouts, R. Madejsky, et al., Astron. Astrophys. 392, 741 (2002).

21. J. T. A. de Jong, B. Yanny, H.-W. Rix, A. E. Dolphin, N. F. Martin, and T. C. Beers, Astrophys. J. 714, 663 (2010).

22. Y. C. Joshi, Mon. Not. R. Astron. Soc. 378, 768 (2007).

23. P. C. van der Kruit, Astron. Astrophys. 157, 230 (1986).

24. L. Lindegren, J. Hernandez, A. Bombrun, et al. (Gaia Collab.), Astron. Astrophys. 616, 2 (2018).

25. J. Maiz-Apell\'aniz, Astron. J. 121, 2737 (2001).

26. P. Mr\'oz, A. Udalski, D. M. Skowron, et al., Astrophys. J. 870, L10 (2019).

27. D. K. Ojha, Mon. Not. R. Astron. Soc. 322, 426 (2001).

28. S. A. Olausen and V. M. Kaspi, Astrophys. J. Suppl. Ser. 212, 6 (2014).

29. A. Pieres, L. Girardi, E. Balbinot, B. Santiago, L. N. da Costa, A. Carnero Rosell, A. B. Pace, K. Bechtol, et al., Mon. Not. R. Astron. Soc. 497, 154 (2020).

30. A. E. Piskunov, N. V. Kharchenko, S. R\"oser, E. Schilbach, and R.-D. Scholz, Astron. Astrophys. 445, 545 (2006).

31. C. Reyl\'e and A. C. Robin, Astron. Astrophys. 373, 886 (2001).

32. H.-W. Rix and J. Bovy, Astron. Astrophys. 21, 61 (2013).

33. A. C. Robin and M. Cr\'ez\'e, Astron. Astrophys. 157, 71 (1986).

34. A. C. Robin, C. Reyl\'e, S. Derri\'ere, and S. Picaud, Astron. Astrophys. 409, 523 (2003).

35. C. K. Rosslowe and P. A. Crowther, Mon. Not. R. Astron.
Soc. 447, 2322 (2015).

36. D. M. Skowron, J. Skowron, P. Mr\'oz, et al., Science
(Washington, DC, U. S.) 365, 478 (2019).

37. M. F. Skrutskie, R. M. Cutri, R. Stiening, M. D. Weinberg, S. Schneider, J. M. Carpenter, C. Beichman, R. Capps, et al., Astron. J. 131, 1163 (2006).

38. A. Udalski, M. K. Szyma\'nski, and G. Szyma\'nski, Acta Astron. 65, 1 (2015).

39. S. Wang, X. Chen, R. de Grijs, et al., Astrophys. J. 852, 78 (2018).

40. M. W. Werner, T. L. Roellig, F. J. Low, G. H. Rieke, M. Rieke, W. F. Hoffmann, E. Young, J. R. Houck, et al., Astrophys. J. Suppl. Ser. 154, 1 (2004).

41. Z. Zheng, C. Flynn, A. Gould, J. N. Bahcall, and S. Salim, Astrophys. J. 555, 393 (2001).
}
\end{document}